\documentclass[11pt]{article}
\usepackage{blois,epsfig}

\def\be{\begin{equation}}
\def\ee{\end{equation}}
\def\bea{\begin{eqnarray}}
\def\eea{\end{eqnarray}}

\begin{document}
\vspace*{4cm}
\title{THE OBSERVED CORRELATIONS FOR THE STRANGE MULTIBARYON  STATES
IN SYSTEMS WITH $\Lambda$-HYPERON FROM pA COLLISION AT MOMENTUM OF
10 GeV/$c$}

\author{ P.Zh.Aslanyan}

\address{Joint Institute for Nuclear Research, Joliot-Curie str.,\\
Dubna, p.o. 141980, Russia}

\maketitle\abstracts{The observed well-known resonances $\Sigma^0$
$\Sigma^{*+}$(1385) and  $K^{*\pm}$(892) from PDG are good tests of
this method.
 Exotic strange multibaryon states have been observed in
the effective mass spectra of: $\Lambda \pi^{\pm}$,$\Lambda \gamma$,
$\Lambda p$, $\Lambda p p$ subsystems. The mean value of mass for
$\Sigma^{*-}(1385)$ resonance is shifted till mass of 1370 MeV/$c^2$
and width is two times larger than the same value from PDG. Such
kind of behavior for width and invariant mass of $\Sigma^{*-}(1385)$
resonance is interpreted as extensive contribution from stopped
$\Xi^-\to\Lambda\pi^-$  and medium effect with invariant mass. The
mean value of mass  for $\Sigma^{*+}(1385)$  from secondary
interactions is also shifted till mass of 1370 MeV/$c^2$. The width
of $\Sigma^0$ is $\approx$ 2 times larger than the experimental
error. There are enhancement production for all observed hyperons.}

\section{Introduction}

Observation of strange multi-baryonic clusters is an exiting
possibility to explore the properties of cold dense baryonic matter
and non-perturbative QCD \cite{jaffe}-\cite{salvini}.

 The experimental data from heavy ion collisions show that the $K^+/\pi^+$ ratio
(\cite{rafel}) is larger at BNL$-$AGS energies than at the highest
CERN$-$SPS energies  and even those at RHIC. The experimental
$\Lambda /\pi^+$ ratio in the $pC$ reaction is approximately two
times larger than this ratio in $pp$ reactions or in $pC$ reactions
within the FRITIOF model at the same energy \cite{pepan09}. However,
there are no sufficient experimental data concerning strange hyperon
production in hadron$-$nucleus and nucleus-nucleus collisions over
the 4$-$50 GeV/$c$ momentum range.

Recently, the existence of discrete nuclear bound states of
$\overline{K}^0$p has been predicted within the phenomenological
Kaonic Nuclear Cluster (KNC) model based on the experimental
information on the $\overline{K}^0$N scattering lengths, kaonic
hydrogen atom, and the $\Lambda^*(1405)$ resonance \cite{knc},
\cite{kienle}.

Although such states were predicted by Wycech \cite{wycech} some
time ago, only recently the availability of experimental facilities
(KEK~\cite{suzuki},\cite{e471}, DISTO~\cite{disto},
FOPI~\cite{fopi}, DAFNE~\cite{finuda}, and OBELIX~\cite{obelix}, in
particular for studying these kind of exotic nuclei, has delivered
first experimental results which triggered a vivid discussion and
project of AMADEUS~\cite{buehler}.

As was shown  \cite{salvini}, the strangeness production in p
annihilation on nuclei and on hydrogen are related to the possible
quark-gluon plasma formation and to the existence of DBKS in dense
hadronic matter.

Following \cite{rafel}, we assume that the above experimental fact
is due to the formation of a 'blob' of QGP.

Experimental evidence for exotic dibaryons in the ($\Lambda$p)
system first came from the observation of S = -1 narrow resonances
by the propane bubble chamber method  at the beam momenta of 7 GeV/c
\cite{shakh} and 10 GeV/c \cite{pom},\cite{hs07},\cite{pepan09}.

\section{($\Lambda, \pi^+$), ($\Lambda, p $) and $\Lambda \gamma$ spectra}

\subsection{($\Lambda, \pi^+$)spectra} \label{sec:sig}

\begin{figure}
\psfig{figure=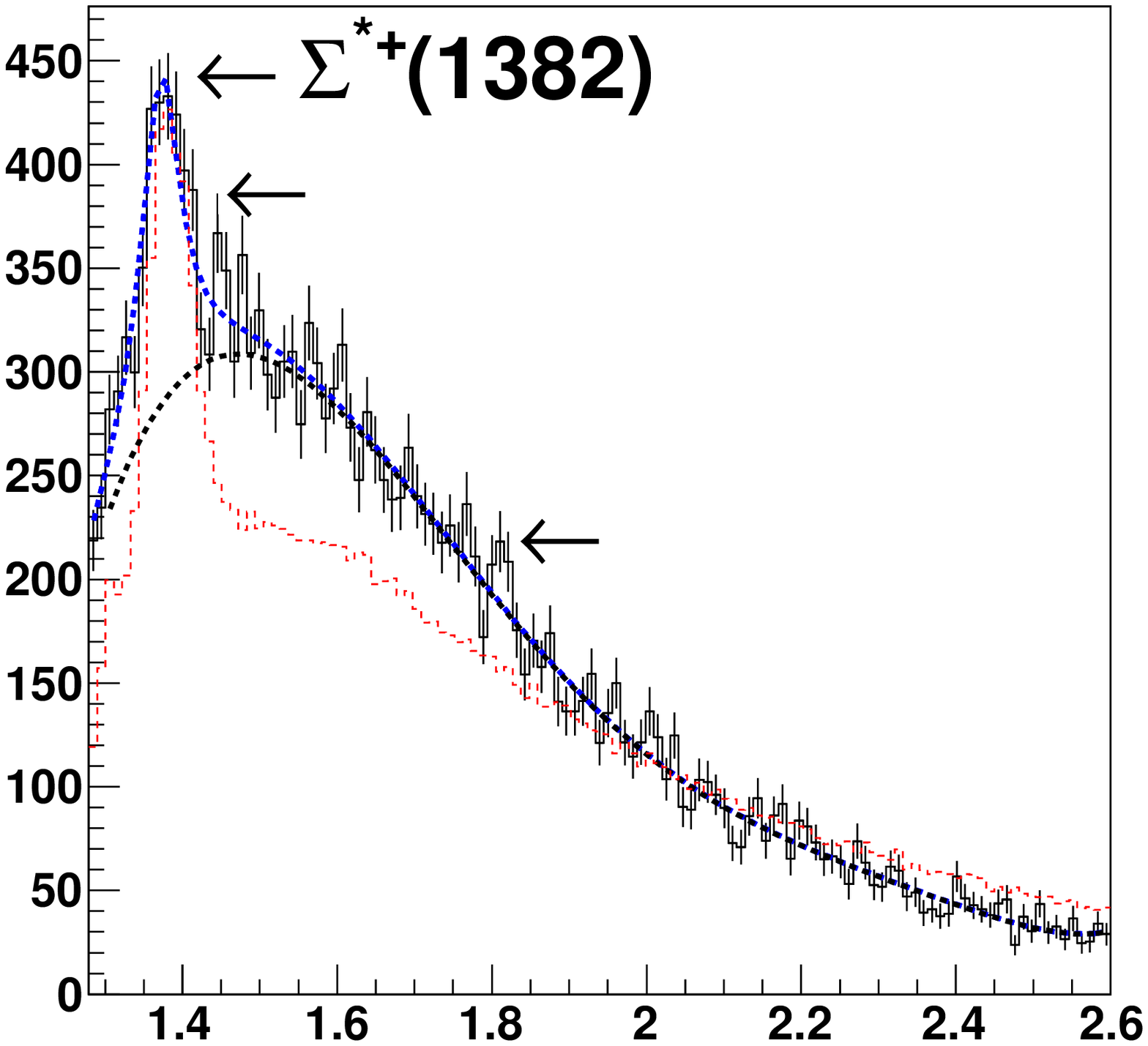,height=1.8in}a)
\psfig{figure=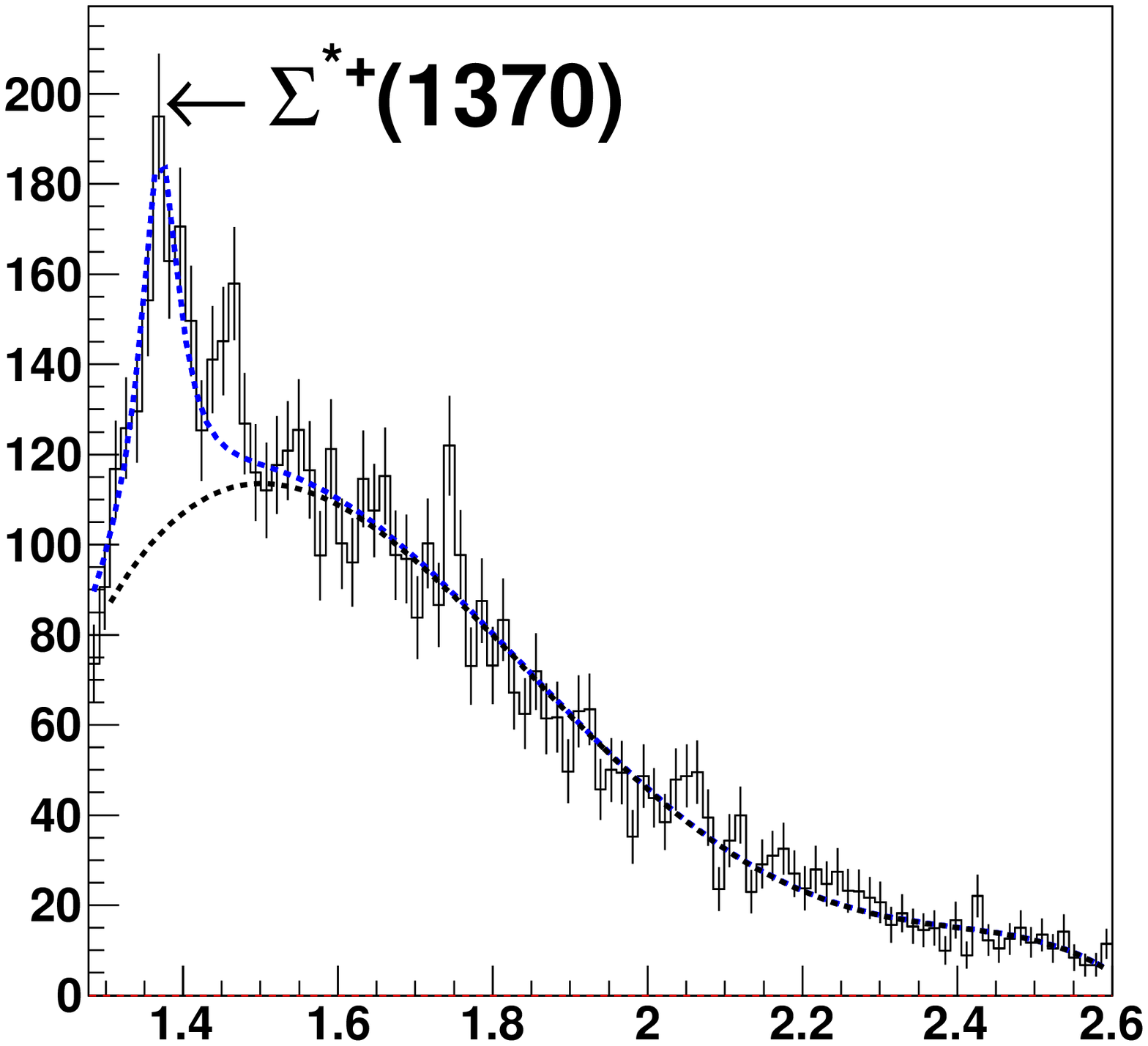,height=1.8in}b)
\psfig{figure=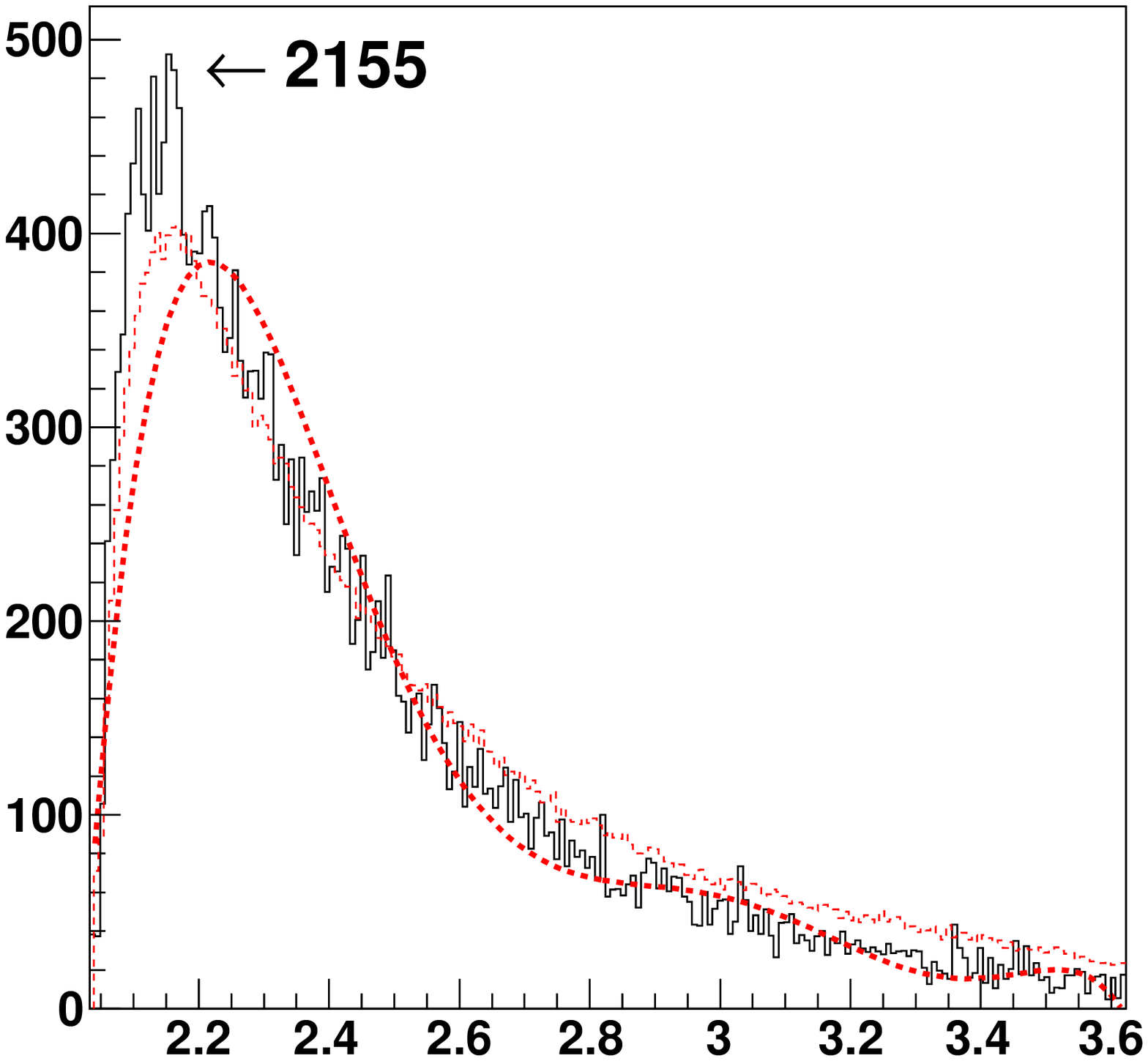,height=1.8in}c)
\psfig{figure=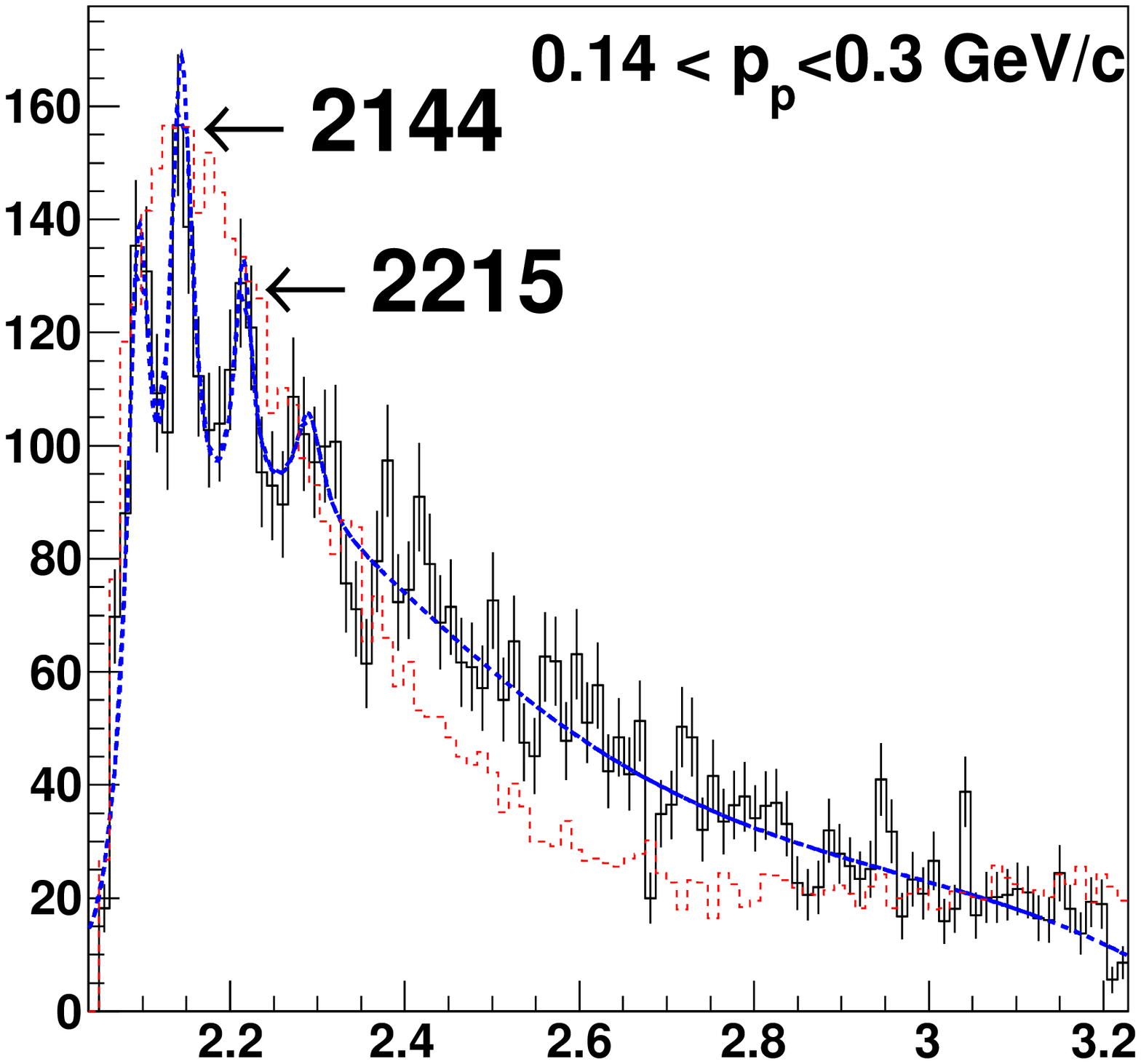,height=1.8in}d)
 \psfig{figure=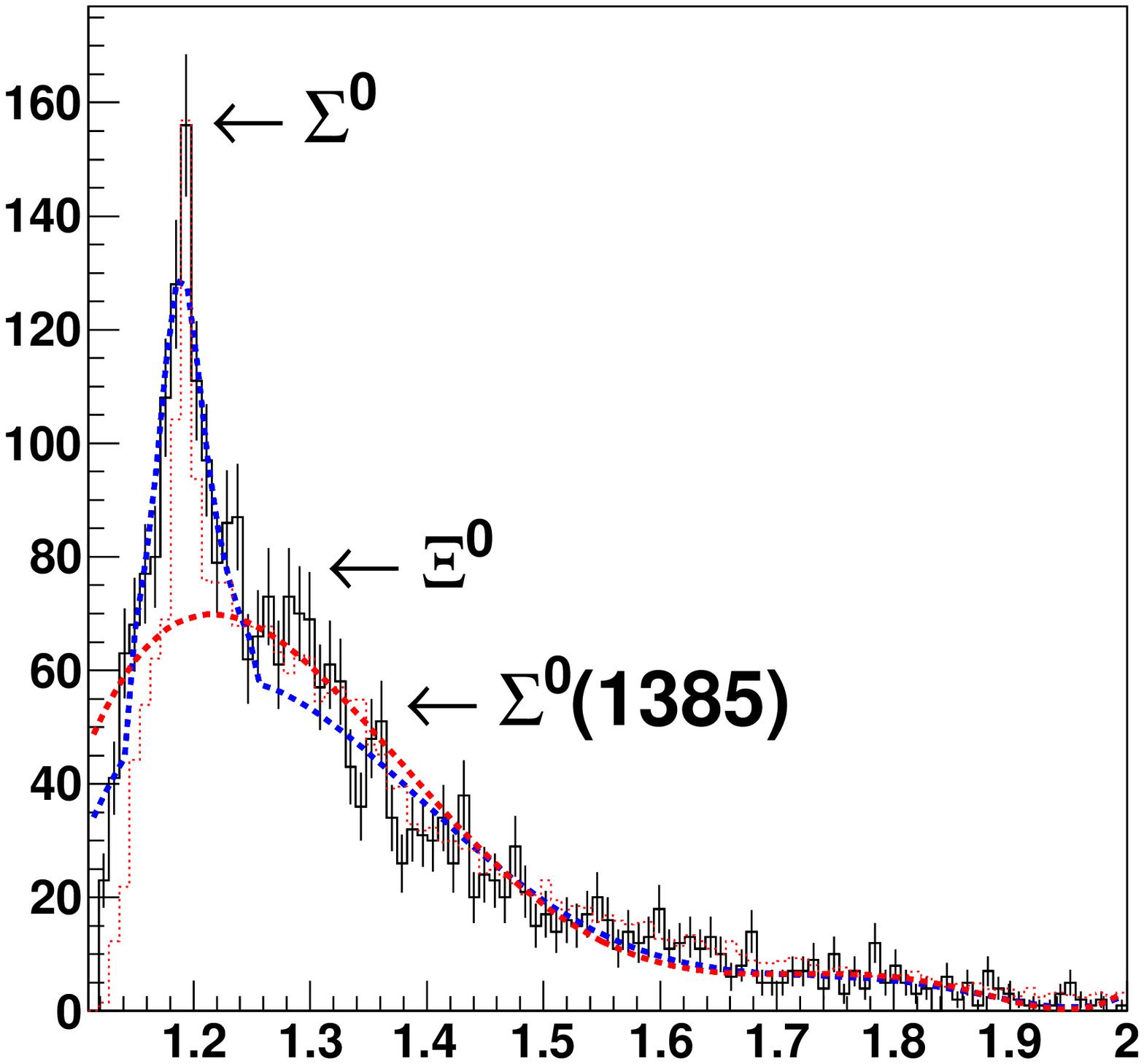,height=1.8in}e)
\psfig{figure=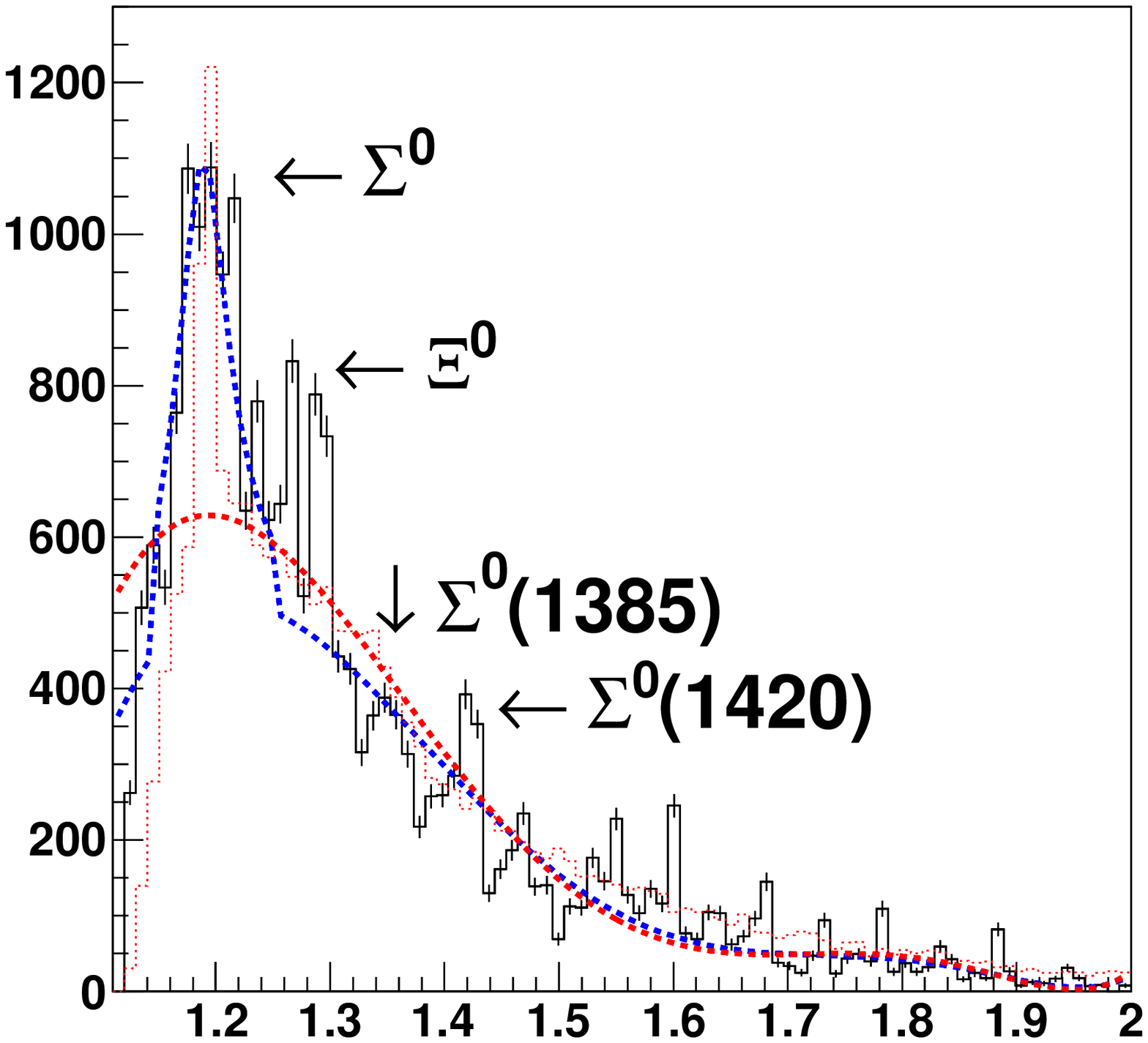,height=1.8in}f)
 \caption{a)The $\Lambda \pi^+$ -
spectrum for all combinations with a bin size of 12 MeV/$c^2$; b)
$\Lambda \pi^+$ spectrum from primary and secondary protons; c)the
$\Lambda p$ - spectrum for all combinations with a bin size of 8
MeV/$c^2$ for p+A interactions from primary protons; d)the $\Lambda
p$ spectrum with stopped protons in the momentum range of
0.14$<P_p<$0.30 GeV/$c$ from primary protons;  e) The $\Lambda
\gamma$ - spectrum for all combinations without total weight from
$\Lambda$ and $\gamma$; f)the $\Lambda\gamma$ spectrum with total
weight. The dashed curve is the experimental background fitted by
polynomial function.\label{lpip}}
\end{figure}

The $\Lambda\pi^+$ effective  mass distribution for all 19534
combinations with a bin size of 17 MeV/$c^2$ at  10 GeV/c is shown
in Fig.\ref{lpip},a \cite{hs07, pepan09}. The dashed histogram  in
the figures is the background simulated by the FRITIOF model. The
mass resolution is $\Delta M/M =0.7$\%, the decay width is $\Gamma
\approx$ 45 MeV/$c^2$. The cross section of $\Sigma^{*+}(1382)$
production is approximately 1.1 mb (600 events in the peak,
13$\sigma$) for the $p+C$ interaction, which is 1.5 times larger
than the estimated cross section. This observed resonance
$\Sigma^{*+}(1382) \to\Lambda \pi^+$ was a good test of this method.
The $\Lambda\pi^+$ effective  mass distribution for total 25199
combinations with a bin size of 12 MeV/$c^2$ from primary at 10
GeV/c and secondary relativistic protons over the momentum $4< P_p<
10$ GeV/c is shown in Fig.\ref{lpip},$b$($\approx$780 events in
peak). The secondary protons is a positive tracks with momentum $4<
P_p< 10$ GeV/c,which are induced interaction. The total contribution
in positive tracks from another $K^+$,$\pi^+$ particles lower than
15 \%. The dashed histogram in Fig.\ref{lpip}, $b$ is the background
by the mixing momentum method. The upper dashed curve
(Fig.~\ref{lpip},$a$ and $b$) is the sum of the background and 1
Breit-Wigner function. The background (lower dashed curve) is the
8th-order polynomial function. The bin size is consistent with the
experimental resolution. There are small signals in mass range of
1450(3.5$\sigma$ and 1750(3.3$\sigma$) MeV/$c^2$. This enhancements
 can interpreted as reflection from $\Lambda^*$ resonances. The mean
value of mass in Fig.\ref{lpip}, $b$ is shifted till 1370 MeV/$c^2$.
The same shift have observed for the mean value of $\Lambda\pi^-$
spectrum \cite{hs07, pepan09}. Such kind of behavior in both case we
can interpreted as medium effect in nucleus from lower energy
$\pi^{\pm}$, because a momentum distributions for $\pi^-$ from the
beam protons and for  $\pi^+$ from secondary protons  have similarly
behavior.

\subsection{($\Lambda, p$) spectra} \label{sec:lp}

In the are published reports \cite{pom}-\cite{pepan09} the ($\Lambda
p$) invariant mass with identified protons is given for the momentum
range of 0.350$< p_p<$ 0.900 GeV/$c$.   There are significant
enhancements in the mass regions of 2100(6.9 $\sigma$),
2175(4.9$\sigma$), and 2285(3.8$\sigma$)MeV/$c^2$. There are also
small peaks at masses of 2225(2.2$\sigma$) and 2353(2.9$\sigma$)
MeV/$c^2$.

Figure \ref{lpip}, $c$  shows the invariant mass for all $\Lambda p$
21500 combinations with bin size 8 MeV/$c^2$ from primary beam
protons.  There is significant signal in region of 2155 MeV/$c^2$(>
6$\sigma$).There are small enhancements in mass regions of 2100,
2212 and 2310 MeV/$c^2$.

Figure \ref{lpip}, $d$  shows the invariant mass of 4669($\Lambda
p$) combinations with a bin size of 14 MeV/$c^2 $ for stopped
protons in the momentum range of 0.14$< p_p<$ 0.30 GeV/$c$. The
dashed curve is the sum of the eight-order polynomial  and 4
Breit-Wigner curves with $\chi^2=30/25$ from fits. There are
significant enhancements in the mass regions of 2100(5.7 $\sigma$),
2150(5.7$\sigma$), 2220(6.1$\sigma$), 2310(3.7$\sigma$), and
2380(3.5$\sigma$)MeV/$c^2$. The significant peak in the mass range
of 2220 MeV/$c^2$ (6.1 S.D.), $B_K$ ~ 120 MeV is confirmed by the
KNC model  prediction \cite{knc} in the $K^- pp \to \Lambda p$
channel.

 The $\Lambda p$ effective mass distribution for
4523 combinations with relativistic protons over a momentum range of
P $>$1.5 GeV/$c$ is presented in \cite{pepan09}, where the events
with the undivided ($\Lambda K^0_s$) are removed. The solid curve is
the 6-order polynomial function ($\chi^2$/n.d.f=271/126). There are
significant enhancements in the mass regions of 2150(4.4 S.D.),
2210(3.8 S.D.), 2270(3.4 S.D.), 2670 (3.1 S.D.), and 2900(3.1 S.D.)
MeV/c$^2$. The observed peaks for the combinations with relativistic
$P>$1.5 GeV/$c$ protons agree with spectra from combination with the
identified protons and with stopped protons.

\subsection{($\Lambda, \gamma$) spectra} \label{sec:lg}

Figure \ref{lpip}, $d$  shows the invariant mass for all $\Lambda
\gamma$ 2904 combinations with bin size 9 MeV/$c^2$ from primary
beam protons and  without total  geometrical efficiency from
$\Lambda$ and $\gamma$. The dashed curve is the sum of the six-order
polynomial and 1 Breit-Wigner functions. The statistical
significance for $\Sigma^0$ is 12$\sigma$(or 220 events in the
peak). There are negligible enhancements in mass range of 1320 and
1380 Mev/$c^2$.

Figure \ref{lpip}, $f$  shows the invariant mass for all $\Lambda
\gamma$ 2904 combinations with bin size 10 MeV/$c^2$ from primary
beam protons and  with total  geometrical efficiency from $\Lambda$
and $\gamma$. The cross section of  production for
$\Sigma^0$($\approx$ 1800 events, $<w_{\gamma}>$=4.1) is equal to
$\sigma$= 3.3 mb  for p+C interaction at 10 GeV/c which is more 2
times larger than simulated cross section by FRITIOF. The observed
width of $\Sigma^0$ is  $\approx$ 2 times larger than value of
experimental errors. There are  small enhancements in mass ranges of
1320 ,1380, 1420, 1550 and 1630 with bin size 10 MeV/$c^2$ which are
interpreted as reflection from enhancement production well known
hyperons.

\section{Conclusion}

A number of important peculiarities were observed in the effective
mass spectrum  \cite{pom}-\cite{pepan09}: $(\Lambda, \pi^{\pm})$,
$(\Lambda, \pi^+ ,\pi^-)$, $(\Lambda,p)$, $(\Lambda, p, p)$,
($\Lambda,\Lambda$), ($\Lambda,p ,\pi^-$), ($\Lambda,K^0_s$),
($K^0_s\pi^{\pm}$) and ($K^0_s p$). There are enhancement signals
from all observed hyperons\cite{pepan09}. The observed width of
$\Sigma^0$ and $\Sigma^-$ is  $\approx$ 2 times larger than PDG
value.

According to A. Gal \cite{gal} and R.Jaffe(proc. J-PARC
workshop-2005), the issue of strange multibaryon states is yet far
from being experimentally resolved and more dedicated, systematic
search is necessary. The search for and study of exotic strange
multibaryon states with $\Lambda$ and $K_s^0$ subsystems at
MPD(NICA, JINR), CBM(FAIR, GSI), p07(JPARC, KEK), OBELIX(CERN) and
AMADEUS(DAFNE, INFN) can  provide  information on their nature and
properties and will be a test for the observed PBC data. Higher
statistics experiments with the mass resolution $\approx$1\% are
needed.

\section*{Acknowledgments}
My thanks "Windows on the World" Organizing Committee and personally
Professor Jean Tran Thanh Van for providing the excellent
organization, warm and stimulating atmosphere during the Conference
and for the financial support.

\end{document}